\documentclass[floatfix,  reprint,
superscriptaddress,
twocolumn]{revtex4-2}

\usepackage{graphicx}
\usepackage{dcolumn}
\usepackage{adjustbox}

\usepackage{float}
\usepackage[utf8]{inputenc}
\usepackage[T1]{fontenc}

\usepackage{bm}
\usepackage{breqn}
\usepackage{xcolor}
\usepackage{ragged2e}
\usepackage{amsmath}

\usepackage[title]{appendix}
\usepackage[normalem]{ulem}

\usepackage{physics}
\usepackage[version=4]{mhchem}
\usepackage{cleveref} 
\crefname{appendix}{Appendix}{Appendices}
\crefname{equation}{Eq.}{Eqs.}
\crefname{figure}{Fig.}{Figs.}
\crefname{table}{Table}{Tables}
\crefname{section}{Section}{Sections}
\crefname{enumi}{case}{cases}
\creflabelformat{appendix}{[#2#1#3]}

\begin{document}

\title{Machine Learning-Guided Discovery of\\ Kagome Superconductors YRu$_3$B$_2$ and LuRu$_3$B$_2$}
\author{Rose Albu Mustaf}
\thanks{These authors contributed equally.}
\affiliation{Department of Physics and Astronomy$,$ Rice University$,$ Houston$,$ Texas 77005$,$ USA}
\affiliation{Rice Center for Quantum Materials$,$ Rice University$,$ Houston$,$ Texas 77005$,$ USA}

\author{Sajilesh K. P.}
\thanks{These authors contributed equally.}
\affiliation{Department of Physics and Astronomy$,$ Rice University$,$ Houston$,$ Texas 77005$,$ USA}
\affiliation{Rice Center for Quantum Materials$,$ Rice University$,$ Houston$,$ Texas 77005$,$ USA}

\author{Sanu Mishra}
\affiliation{Department of Physics and Astronomy$,$ Rice University$,$ Houston$,$ Texas 77005$,$ USA}
\affiliation{Rice Center for Quantum Materials$,$ Rice University$,$ Houston$,$ Texas 77005$,$ USA}

\author{Junze Deng}
\affiliation{Department of Applied Physics$,$ Aalto University School of Science$,$ FI-00076 Aalto$,$ Finland}

\author{Yi Jiang}
\affiliation{Donostia International Physics Center (DIPC)$,$ Paseo Manuel de Lardiz\'{a}bal.~20018$,$ San Sebasti\'{a}n$,$ Spain}

\author{Kaja H.~Hiorth}
\affiliation{Department of Applied Physics$,$ Aalto University School of Science$,$ FI-00076 Aalto$,$ Finland}

\author{Eeli O.~Lamponen}
\affiliation{Department of Applied Physics$,$ Aalto University School of Science$,$ FI-00076 Aalto$,$ Finland}

\author{Martin Gutierrez-Amigo}
\affiliation{Department of Applied Physics$,$ Aalto University School of Science$,$ FI-00076 Aalto$,$ Finland}

\author{P\"aivi T\"orm\"a} 
\affiliation{Department of Applied Physics$,$ Aalto University School of Science$,$ FI-00076 Aalto$,$ Finland}

\author{Miguel A.L. Marques}
\affiliation{
 Research Center Future Energy Materials and Systems of the University Alliance Ruhr and Interdisciplinary Centre for Advanced Materials Simulation$,$ Ruhr University Bochum$,$ Universitätsstraße 150$,$ D-44801 Bochum$,$ Germany}

\author{B. Andrei Bernevig}
\affiliation{Donostia International Physics Center (DIPC)$,$ Paseo Manuel de Lardiz\'{a}bal.~20018$,$ San Sebasti\'{a}n$,$ Spain}
\affiliation{Department of Physics$,$ Princeton University$,$ Princeton$,$ NJ 08544$,$ USA}
\affiliation{IKERBASQUE$,$ Basque Foundation for Science$,$ 48013 Bilbao$,$ Spain}

\author{Emilia Morosan}
\email[corresponding author: E. Morosan: ]{emorosan@rice.edu}
\affiliation{Department of Physics and Astronomy$,$ Rice University$,$ Houston$,$ Texas 77005$,$ USA}
\affiliation{Rice Center for Quantum Materials$,$ Rice University$,$ Houston$,$ Texas 77005$,$ USA}

\date{\today}

\begin{abstract}
We report the experimental discovery of bulk superconductivity in two kagome lattice compounds, YRu$_3$B$_2$ and LuRu$_3$B$_2$, which were predicted through machine learning-accelerated high-throughput screening combined with first principles calculations. These materials crystallize in the hexagonal CeCo$_3$B$_2$-type structure with planar kagome networks formed by Ru atoms. We observe superconducting critical temperatures of $T_{c} = 0.81$~K for YRu$_3$B$_2$ and $T_{c} = 0.95$~K for LuRu$_3$B$_2$, confirmed through magnetization, specific heat, and electrical transport measurements. Both compounds exhibit nearly 100\% superconducting volume fractions, demonstrating bulk superconductivity. Compared with isostructural LaRu$_3$Si$_2$, YRu$_3$B$_2$ and LuRu$_3$B$_2$ show a more dispersive Ru local $d_{x^2-y^2}$ quasi-flat band (and thus a reduced DOS at $E_F$) together with an overall hardening of the phonon spectrum, both of which lower the electron-phonon coupling (EPC) constant $\lambda$. Meanwhile, the dominant real-space EPC between Ru local $d_{x^2-y^2}$ states and the low-frequency Ru in-plane local $x$ branch remains nearly unchanged, indicating that the reduction of $\lambda$ originates from the $d_{x^2-y^2}$ DOS reduction and the overall phonon hardening. Superfluid weight calculations show that conventional contributions dominate over quantum geometric effects due to the dispersive nature of bands near the Fermi level. This work demonstrates the effectiveness of integrating machine learning screening, first principles theory, and experimental synthesis for accelerating the discovery of new superconducting materials.
\end{abstract}

\maketitle
\section{Introduction}
During the century since superconductivity was discovered experimentally in Hg~\cite{Onnes}, serendipity has been the predominant way of finding new superconductors. Even after the theory of conventional superconductivity had been established by Bardeen, Cooper, and Schrieffer (BCS)~\cite{BCStheory}, the new understanding of the role of electron-phonon correlations aided little in predicting novel superconducting materials and their critical parameters. New classes of superconductors, such as cuprates~\cite{cuprates}, pnictides~\cite{pnictides}, heavy fermion superconductors~\cite{HFSC}, organic superconductors~\cite{organicSC}, and very recently nickelates~\cite{nickelates} or even the conventional superconductor MgB$_2$~\cite{Danfeng}, have been the result of accidental discoveries. One success of the theoretical prediction of superconductivity has been metallic hydrogen~\cite{ashcroft1968metallic} and hydrides~\cite{ashcroft2004hydrogen}, which were conjectured to be the best candidates for high temperature superconductivity, based on the strong bonds and the light mass of hydrogen. Remarkably, hydrides currently hold the record for the highest experimentally observed superconducting critical temperatures~\cite{Roadmap_RTS}, although at extremely high pressures. These materials were theoretically predicted using first-principles calculations before their experimental synthesis and characterization. The successful prediction and subsequent verification of superconductivity in these hydrogen-rich compounds represents a paradigm shift in materials discovery, demonstrating the power of computational methods to guide experimental efforts toward new superconductors.

Despite the success of first-principles theories in understanding the properties of superconducting materials, the immense vastness of the materials space and the high computational expense of these calculations impose a fundamental bottleneck for materials discovery. To address this challenge, machine learning (ML) has recently emerged as a powerful approach capable of accelerating the prediction of new superconductors~\cite{Miguel2025,Hennig2025,DengLaRuSi2025,2ndHennig2025,LesserML2025}. ML models offer a means to rapidly screen candidate materials, which can then be strategically combined with conceptual guidelines derived from established theory, such as the identification of specific lattice motifs or structural features, for targeted searches. This ML-accelerated, theory-guided approach can be used to predict superconductivity in promising classes of materials, thereby guiding subsequent experimental synthesis and investigation.

In this context, materials with kagome lattices host quasi-flat bands with a large density of states, and have thus been a promising candidate for superconductivity. Flat bands with attractive interactions need quantum geometry~\cite{QGreview2025} to enable supercurrent in flat bands~\cite{peotta:2015,Torma2022}, while with repulsive interactions, partially filled flat bands predominantly magnetize~\cite{huvafekhaulebernevig}. Recently, we conducted a comprehensive, ab initio investigation into the superconductivity of the flat-band kagome metal LaRu$_3$Si$_2$~\cite{DengLaRuSi2025}. Furthermore, we performed an ML-accelerated high-throughput screening of the entire 1:3:2 kagome family and identified several stable materials predicted to exhibit superconductivity.

Here, we report the first successful experimental confirmation of bulk superconductivity in one of these predicted systems: YRu$_3$B$_2$. Furthermore, we report a second isostructural, new superconductor, LuRu$_3$B$_2$. The Lu compound was initially overlooked in our high-throughput predictions~\cite{DengLaRuSi2025} because its calculated phonon spectrum exhibits weakly imaginary modes, indicating a possible low-temperature dynamical instability.

Although the predicted critical temperatures ($T_c$) are higher (3.37~K for YRu$_3$B$_2$ and 1.88~K for LuRu$_3$B$_2$) than the experimental observations (0.81 K and 0.95 K, respectively), this result clearly demonstrates the merit of combining ML-driven screening with refined, post-screening ab initio calculations and experiments. This integrated approach provides a rigorous and efficient pathway from theoretical prediction to experimental realization of novel superconducting materials.

 
\section{Experiments}
\subsection{Methods}\label{ExpMeth}
\textit{R}Ru$_3$B$_2$ (\textit{R} = Y, Lu) samples were prepared from high-purity elements (Y 99.99\%, Lu 99.99\%, Ru 99.99\%, and B 99.99\%) in a stoichiometric composition 1:3:2. The elements were arc-melted on a water-cooled copper hearth in an argon atmosphere and remelted several times to ensure homogeneity. The mass loss during the melting process was negligible. The crystal structure was confirmed using powder X-ray diffraction in a Bruker D8 Advance diffractometer with Cu K$\alpha$ radiation. Rietveld refinements were performed using FullProf suite software \cite{Fullprof}. Magnetization measurements M(T;H) were performed in a QD Magnetic Property Measurement System (MPMS) equipped with the iQuantum Helium-3 option. Specific heat C${_p}$(T;H) measurements were performed in a Quantum Design (QD) Dynacool Physical Property Measurement System (PPMS) with a dilution refrigerator (DR) option.  


\subsection{Results and Discussion} \label{Results}

\begin{figure}
    \centering
    \includegraphics[width = \linewidth]{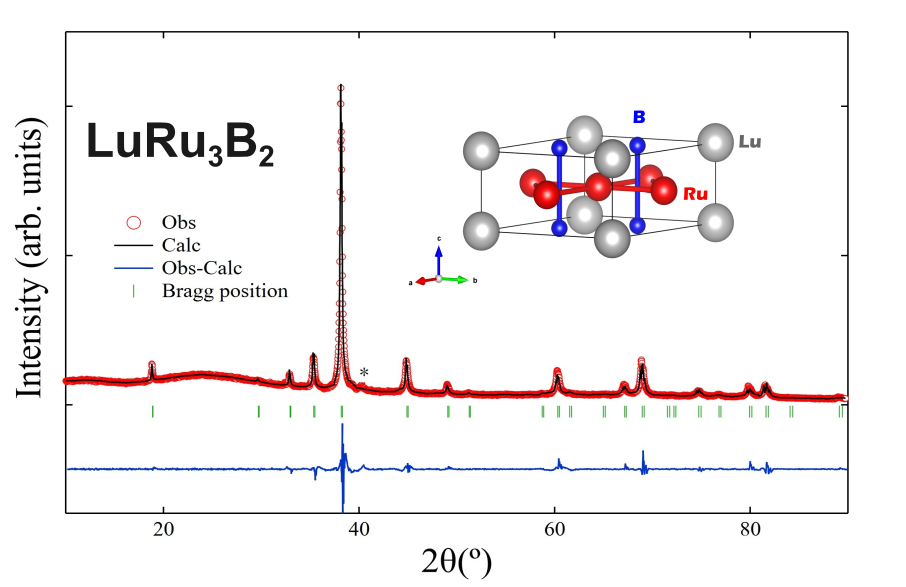}
    \caption {Powder X-ray pattern for LuRu$_3$$B_2$ (red), with calculated pattern (black) and Bragg peak positions (green ticks) for space group P6/mmm. The asterisk denotes residual Ru. Inset: crystallographic unit cell of \textit{R}Ru$_3$B$_2$.}
    \label{fig:xray}
\end{figure}

\textit{R}Ru$_3$B$_2$ (\textit{R} = Y, Lu) form in the CeCo$_3$B$_2$-type structure, which crystallizes in the hexagonal P6/mmm space group (\# 191). In this structure, the Ru atoms arrange into a planar kagome network in the $ab$ plane. The unit cell of \textit{R}Ru$_3$B$_2$ is shown in the inset of Fig. \ref{fig:xray}. Powder X-ray diffraction confirms the crystal structure and phase purity of the synthesized samples. The measured patterns (shown in Fig. \ref{fig:xray} for LuRu$_3$B$_2$) can be fit to the expected CeCo$_3$B$_2$ structure, with the exception of a minute peak corresponding to elemental Ru, marked by an asterisk. Rietveld refinement of the patterns for the two compounds yields lattice parameters of $a$ = 5.475 \AA~and $c$ = 3.027 \AA~for YRu$_3$B$_2$ and  $a$ = 5.448 \AA~and $c$ = 3.014 \AA~for LuRu$_3$B$_2$. 

\begin{figure} [H]
\includegraphics[width=1 \linewidth]{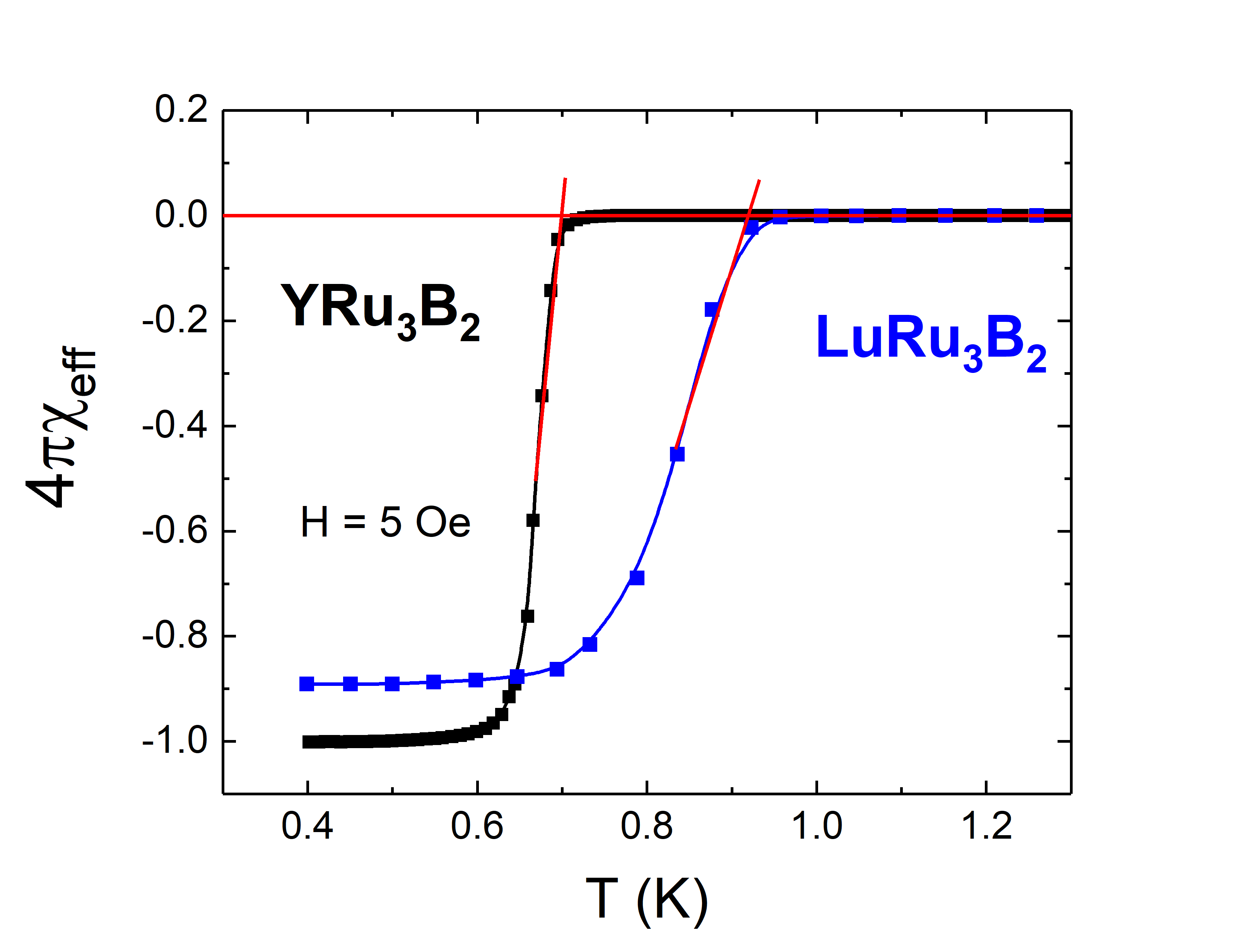}
    \caption{Magnetic susceptibility $4 \pi \chi$$_{eff}$ as a function of temperature for YRu$_3$B$_2$ (black) and LuRu$_3$B$_2$ (blue) at H = 5 Oe.}
    \label{fig:Chi(T)}
\end{figure}

\begin{figure}
\includegraphics[width=1 \linewidth]{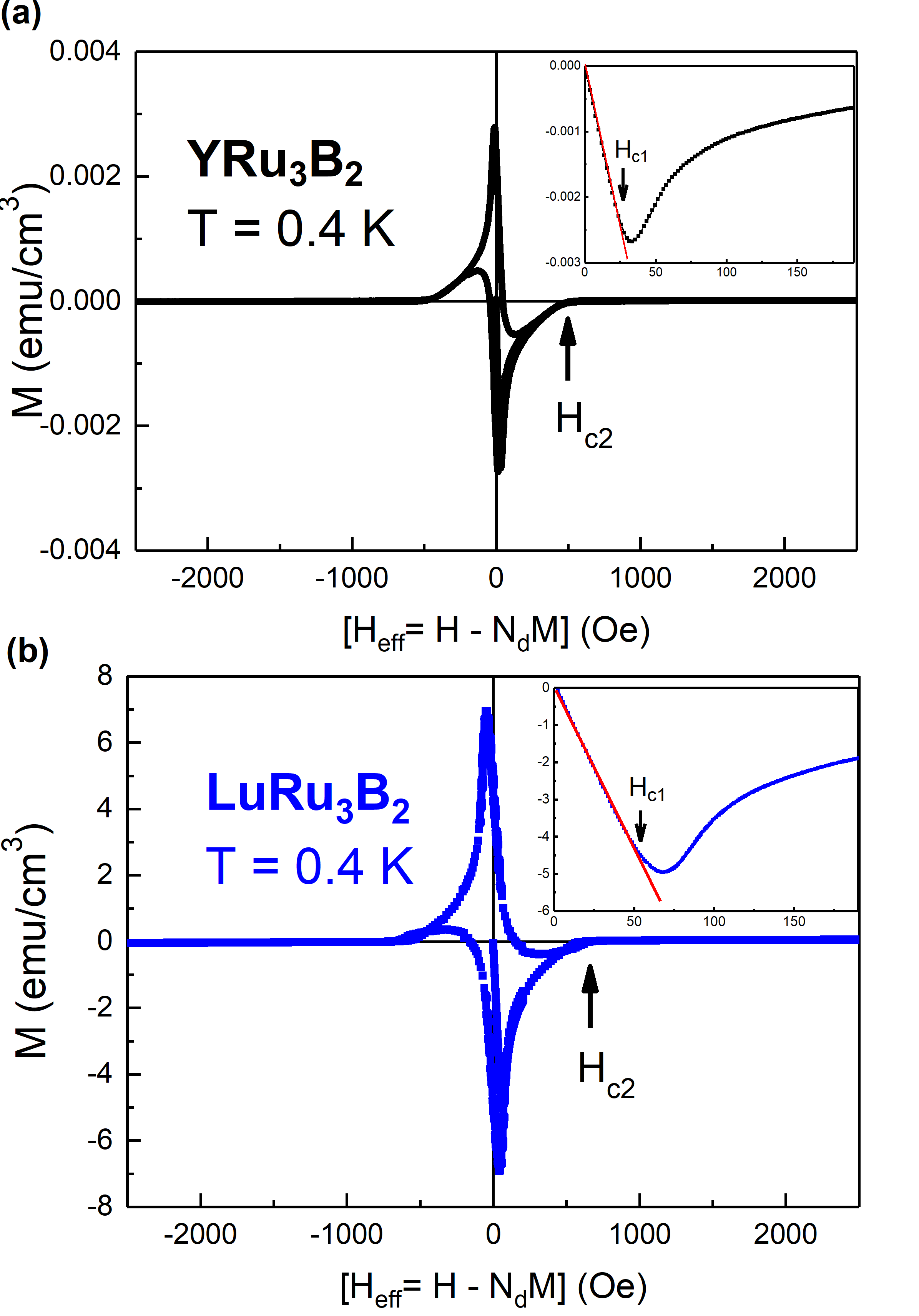}
    \caption{ Magnetization for (a) YRu$_3$B$_2$ (b) LuRu$_3$B$_2$ as a function of H$_{eff}$ at T = 0.4 K, where H$_{eff}$ = H - N$_d$ M. Inset: linear fit of the low H magnetization showing where M(H) deviates from linearity, giving an estimate for H$_{c1}$.}
    \label{fig:MH}
\end{figure}

\begin{figure}
\includegraphics[width= \linewidth]{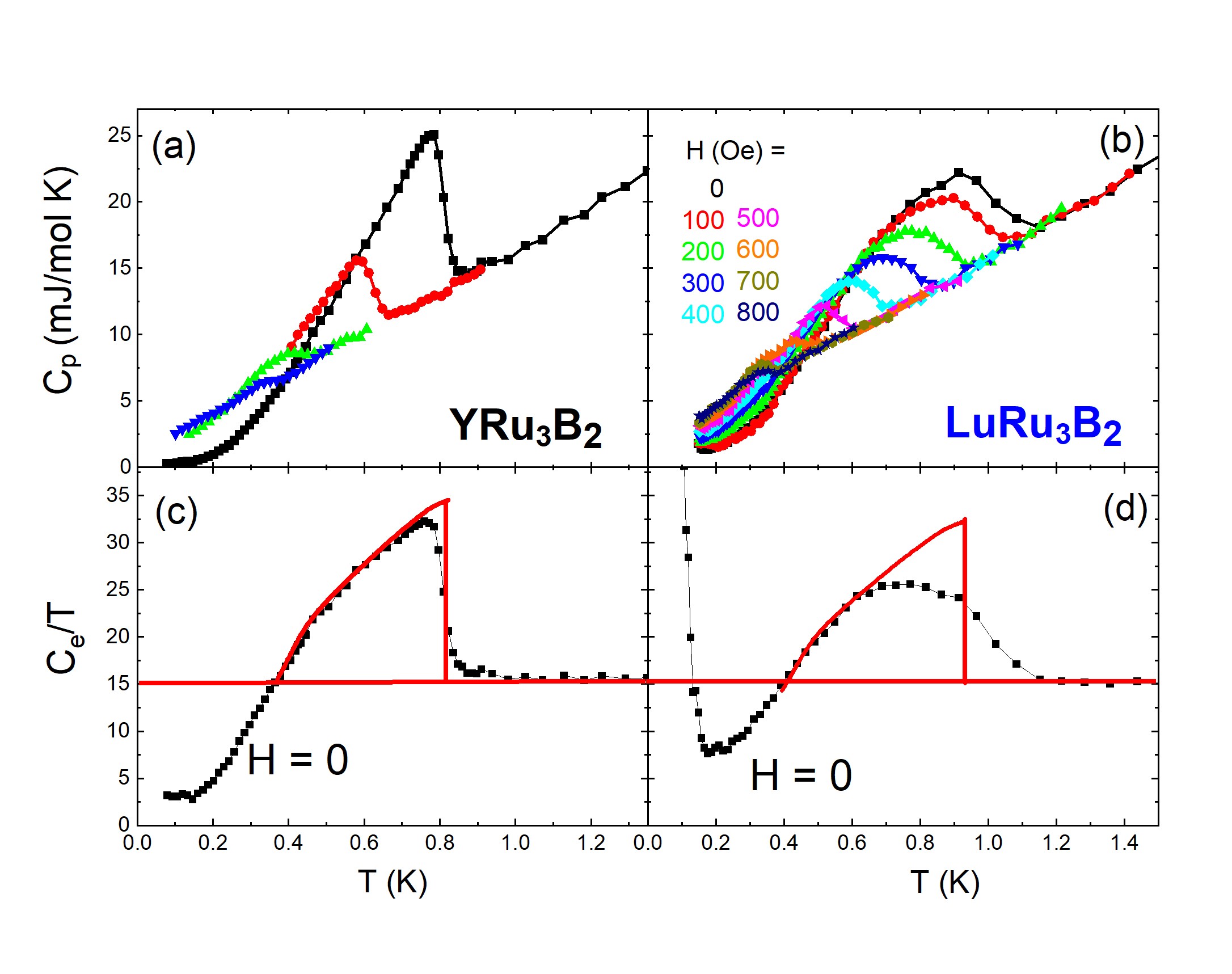}
    \caption{C$_p$(T) data for (a) YRu$_3$B$_2$ and (b) LuRu$_3$B$_2$ in applied fields ranging from 0-800 Oe. }
    \label{fig:HC}
\end{figure}

\begin{figure}
\includegraphics[width= \linewidth]{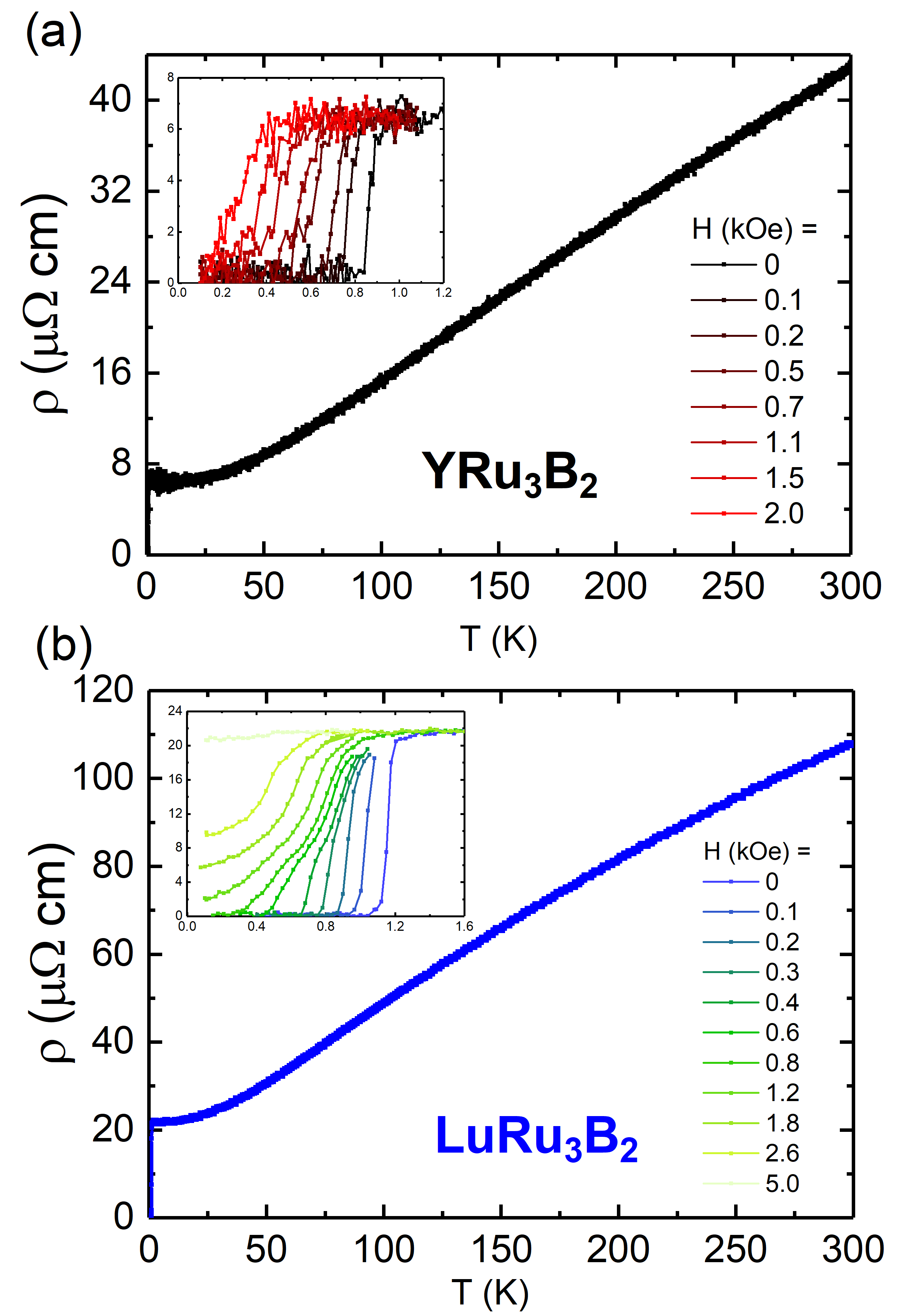}
    \caption{$\rho$(T) data for (a) YRu$_3$B$_2$ and (b) LuRu$_3$B$_2$ at zero field in the range 50 mK-300 K. The insets show the superconducting transitions captured at different applied fields.}
    \label{fig:rhoT}
\end{figure}

We confirmed experimentally the superconductivity in YRu$_3$B$_2$ (black) and LuRu$_3$B$_2$ (blue) with thermodynamic and transport measurements (Figs. \ref{fig:Chi(T)}-\ref{fig:rhoT}). The measured susceptibility data $\chi = M/H$ was scaled by 4$\pi$ and corrected for demagnetizing effects, where $4\pi \chi_{eff} = 4\pi \chi/(1-N_d \chi)$. A demagnetizing factor $N_d = 1/3$ was used for a nearly spherical sample morphology. The resulting superconducting fraction is close to 100\% and 90\% respectively in YRu$_3$B$_2$ and LuRu$_3$B$_2$, confirming the bulk superconducting state in both compounds. For an applied field H = 5 Oe, the critical temperatures, determined by the onset of diamagnetism in \cref{fig:Chi(T)}, are T$_c$ = 0.70 K (YRu$_3$B$_2$) and 0.93 K (LuRu$_3$B$_2$). The critical field values can be estimated from the low temperature magnetization isotherms M(H) (T = 0.4 K) shown in \cref{fig:MH}. A linear fit of the low H magnetization gives an estimate of the lower critical field H$_{c1}$, taken as the H value where M(H) deviates from linearity (vertical arrow, inset). H$_{c1}$ is close to 35 Oe and 48 Oe for YRu$_3$B$_2$ (black) and LuRu$_3$B$_2$ (blue) compounds, respectively. In the T = 0 limit, H$_{c1}(0)$ can be estimated for both compounds, using \cite{Onnes}:
\begin{equation}
   H_{c1}(T) = H_{c1}(0)  [1-(T/T_{c})^2]
\end{equation}
This gives H$_{c1}(0)$ = 52 Oe and 59 Oe for for YRu$_3$B$_2$ and LuRu$_3$B$_2$, respectively. The upper critical field H$_{c2}$ is determined as the field where M(H) approaches zero (vertical arrows, Fig. \ref{fig:MH}). At T = 0.4 K, H$_{c2}$ is close to 530 Oe for YRu$_3$B$_2$ and close to 660 Oe for LuRu$_3$B$_2$. Both of these values are consistent with the findings from specific heat (discussed below). 

We were able to trace the superconducting state to lower temperatures and higher fields in specific heat (Fig. \ref{fig:HC}) measured down to T = 60 mK and applied magnetic field up to 800 Oe. A sharp jump was observed in the H = 0 specific heat at T$_c$ = 0.81 K and 0.95 K, respectively, for YRu$_3$B$_2$ and LuRu$_3$B$_2$. Given that magnetic susceptibility was measured in a small applied field, the specific heat T$_c$ values of H = 0 are slightly higher because, as expected, the increasing applied magnetic field suppresses T$_c$ (fig. \ref{fig:HC}). The normal-state specific heat data for both samples can be modeled by the relation: $C_p/T = \gamma_{n} + \beta T^{2}$, where \(\gamma_{n}\) and \(\beta\) represent the electronic and phononic coefficients, respectively. The fitting parameters for H = 0 are determined as \(\gamma_{n}= 15.1\, \mathrm{mJ/mol K}^2\), \(\beta = 0.16 \, \mathrm{mJ/mol K}^4\) for the YRu$_3$B$_2$ sample, and~\(\gamma_{n} = 15.2 \ \mathrm{mJ/mol K}^2\), \(\beta = 0.48 \, \mathrm{mJ/mol K}^4\) for the LuRu$_3$B$_2$ sample. The superconducting electronic specific heat coefficient $\gamma_e$ can be estimated using $\gamma_n$ and the residual electronic specific heat coefficient $\gamma_{res}$, with the latter estimated from C$_e/T$ at T = 60 mK and H = 0 (Fig. \ref{fig:HC}). Given that $\gamma_{res}$ is much smaller than $\gamma_n$ for both compounds, we approximate the electronic specific heat coefficient $\gamma_{e}$ = $\gamma_{n} - \gamma_{res} \sim \gamma_{n}$ for both YRu$_3$B$_2$ and LuRu$_3$B$_2$. 

The H = 0 entropy-conservation construct shown in Fig. \ref{fig:HC}(c, d) for YRu$_3$B$_2$ and LuRu$_3$B$_2$, respectively, produces values for the jump in the electronic specific heat C$_e$ at $T_c$, $\Delta C_e/\gamma T_c \sim$ 1.1 and 1.26. This value is close to, albeit slightly lower than the BCS predicted value of 1.43. This deviation can arise for a number of reasons, such as grain boundary effects (expected especially in polycrystalline samples) and non-BCS gap features, such as anisotropic or nodal superconducting gap. Our mean-field calculation based on a tight-binding model constructed from DFT band structure calculation gives pairing gaps non-uniform among the orbitals, as described in Section \ref{MFtightbinding}. An upturn in the C$_e$/T for the Lu compound (Fig. \ref{fig:HC}d) is likely due to a nuclear Schottky anomaly at the low temperatures.

\begin{table*}[t]
\caption{Summary of parameters describing YRu$_3$B$_2$ and LuRu$_3$B$_2$ properties.}
\label{tab:summary}
\centering

\begin{ruledtabular}

\begin{tabular}{lcccccccccc}

Compound
& $T_c$ & $H_{c1}(0)$, $H_{c2}(0)$ & $\gamma_n$ & $\beta$ & $\Delta C_e/\gamma_n T_c$ &  $\lambda^{exp}_{e-ph}$ & $m^*$ & $\lambda^{exp}_L(0)$ & $\lambda^{th}_L(0)$ \\
& (K) & (Oe) & (mJ/molK$^{2}$) & (mJ/molK$^{4}$) &  &  & ($m_e$) & (nm) & (nm) \\
\hline
YRu$_3$B$_2$ & 0.81 & 52, 1000 & 15.1 & 0.16 & 1.1 & 0.44 & 1.44 & 32.6 & 32 \\
LuRu$_3$B$_2$ & 0.95 & 59, 867 & 15.2 & 0.48 & 1.26 & 0.41 & 1.41 & 32 & 36 \\
\end{tabular} 

\end{ruledtabular}

\end{table*}

\noindent Using the experimental \(\beta\) values, along with the universal gas constant \(R\), and considering \(N\) = 6 atoms per unit cell, the Debye temperature (\(\theta_{D}\)) can be calculated as: $\theta_{D} = \left( 12 \pi^{4} R N / 5 \beta \right)^{1/3}$. The Debye temperature turns out to be 283.7 K for YRu$_3$B$_2$ and 453 K for the LuRu$_3$B$_2$ compound. The electron-phonon coupling constant in these materials is determined using the following relation:

\begin{equation}
    \lambda^{exp}_{e-ph} = \frac{1.04 + \mu^*\ln(\theta_D/1.45T_c)}{(1 - 0.62\mu^*)\ln(\theta_D/1.45T_c) - 1.04}
\end{equation}

\begin{figure}
\includegraphics[width=1 \linewidth]{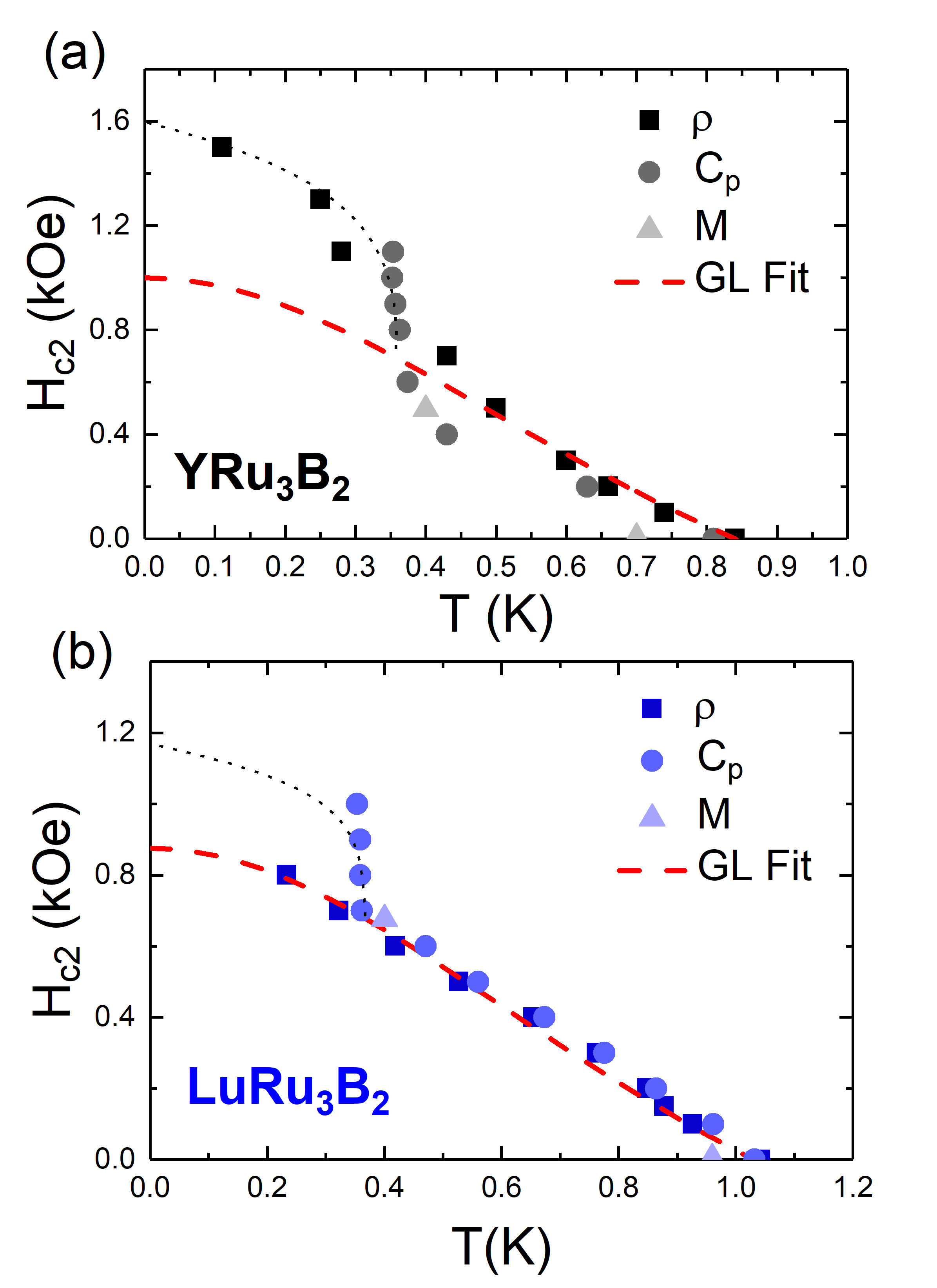}
    \caption{ H$_{c2}$(T) phase diagram for (a) YRu$_3$B$_2$ and (b) LuRu$_3$B$_2$, with the GinzburGLandau (GL) fit (\cref{eq:hc2}) shown in red. }
    \label{fig:H-T}
\end{figure}
Here $\mu^*$ is the repulsive Coulomb interaction term, which is generally considered 0.13 for intermetallic samples. Substituting the values of $\theta_D$ and $T_{c}$ yields \(\lambda^{exp}_{e-ph} \) = 0.44 (for YRu$_3$B$_2$) and 0.41 (for LuRu$_3$B$_2$), indicating the weakly coupled nature of superconductivity in both compounds. The electron-phonon coupling obtained from DFT calculations (Section \ref{ElPhCalc}) is $\lambda^{th}_{e-ph} =$ 0.477 and 0.561 for the Y and Lu compounds, respectively, very close to the estimates from the experiment.

\noindent The Sommerfeld specific heat coefficient $\gamma$ allows us to estimate the ideal London penetration depth $\lambda^{exp}_L(0)$: $\lambda_L(0) = (m^*/\mu_0ne^2)^{1/2}$ , where $m^*$ is the electron effective mass scaled by the bare electron mass $m_e$, $m^*/m_e = (1+\lambda^{exp}_{e-ph}$) and n is the carrier concentration. Due to the absence of single crystals that are required for Hall measurements and therefore experimental n estimates, we calculate the carrier concentration as $n = 3/V$ with V the unit cell volume determined experimentally from X-ray diffraction refinements. The value in the numerator here corresponds to the 3 electrons contributed by the trivalent ions Y$^{3+}$ and Lu$^{3+}$ in the two compounds. This yields $n~=$ 3.82$*10^{-2}$ \AA$^{-3}$ and 3.87$*10^{-2}$ \AA$^{-3}$ for YRu$_3$B$_2$ and LuRu$_3$B$_2$, respectively. The effective masses are $m^* = $ 1.44 and 1.41 m$_e$ for the Y and Lu compounds, respectively. Using these estimates, we can determine $\lambda^{exp}_L(0)$ = 32.6 nm (Y) and 32 nm (Lu). 
We have calculated the superfluid weight (superfluid stiffness) from DFT data (described in Section \ref{SFWCalc}) and obtained the theoretical London penetration depth $\lambda^{th}_L(0)$ = 32 nm (Y) and 36 nm (Lu) in the xy-plane, and these  are in good agreement with the experimental values. The superconducting properties for YRu$_3$B$_2$ and LuRu$_3$B$_2$ compounds are summarized in Table I.

The electrical resistivity measurements show metallic behavior for both samples, with  residual resistivity ratios RRR = $\rho(300K)/\rho_0$ of 6.8 and 5 respectively for YRu$_3$B$_2$ and LuRu$_3$B$_2$ (Fig. \ref{fig:rhoT}). In the low temperature range, the superconducting transition is observed, with the H = 0 resistivity reaching 0 around 0.84 K (YRu$_3$B$_2$) and 1.04 K (LuRu$_3$B$_2$), coinciding with thermodynamic measurements. The insets in Fig. \ref{fig:rhoT} show the expected suppression of the superconducting temperature with applied field. The superconducting transition at each field is defined by the temperature at which the resistance reaches zero. From transport measurements, we can determine the residual resistivity $\rho_0$ to estimate the mean free path $l_0$ for both compounds:
\begin{equation}
    l_0=\frac{3\pi^2\hbar^3}{e^2m^{*2}\rho_0v_F^2}
\end{equation}
where the Fermi velocity $v_F$ is calculated from 
\begin{equation}
    v_F= \frac{\hbar}{m^*}(3\pi^2n)^{1/3}
\end{equation} 

\noindent This gives a mean free path $l_0$= 18.5 nm for  YRu$_3$B$_2$ and 5.2 nm for LuRu$_3$B$_2$. Fig. \ref{fig:H-T} shows the H$_{c2}$ - T phase diagram for YRu$_3$B$_2$ (panel (a), black) and LuRu$_3$B$_2$ (panel (b), blue),  constructed from resistivity (squares), heat capacity (circles) and magnetization (triangles) measurements. The H$_{c2}(0)$ critical field values for the two compounds are determined from the Ginzburg-Landau (GL) fits \cite{Onnes} (red dashed lines): 

\begin{equation}
H_{c2}(T)
=
H_{c2}(0)\,
\frac{1 - \left( T/T_c \right)^2}
{1 + \left( T/T_c \right)^2}
\label{eq:hc2}
\end{equation} 

 \noindent The resulting $H_{c2}$(0) values are close to 1 kOe and 0.87 kOe for YRu$_3$B$_2$ and LuRu$_3$B$_2$, respectively. 
It appears that the superconducting state cannot be described by a simple GL picture, as the $H_{c2} - T$ phase diagram deviates from the conventional predicted behavior in both compounds. The black dotted line at low temperatures is a guide to the eye and gives approximate critical field values around 1.6 kOe and 1.2 kOe at $T$ = 0 K for YRu$_3$B$_2$ and LuRu$_3$B$_2$, respectively.
Similar $H_{c2}-T$ phase diagrams have previously been reported in YbSb$_2$ \cite{PhysRevB.85.214526}, KFe$_2$As$_2$ \cite{PhysRevLett.119.217002}, single-layer NbSe$_2$ \cite{PhysRevB.99.220501},  UPt$_3$ \cite{UPt3}, UTe$_2$  \cite{UTe2, Rosuel} and other heavy fermion superconductors, and have been attributed to a number of possible effects: multiband or multigap superconductivity, a superconducting state with a multi-component order parameter, intrinsic structural or electronic inhomogeneity, or internal strain. Resolving the exact origin of the $H_{c2}$ behavior at low temperatures 
in YRu$_3$B$_2$ and LuRu$_3$B$_2$ is left to a future study based on detailed theoretical calculations and experimental probes (including AC susceptibility, $\mu$SR, optical measurements, STM).

 \noindent We estimated the effective coherence length $\xi_{eff}(0)$ for both compounds using: 
\begin{equation}
    H_{c2}(0) = \frac{\Phi_0}{2 \pi \xi_{eff}(0)^2} 
\end{equation}
where the flux quantum $\Phi_0~ =$ \(2.07\times 10^{-15}\) Wb \cite{Onnes}. This gives  $\xi_{eff}(0) =$ 57.3 nm and 61.6 nm for YRu$_3$B$_2$ and LuRu$_3$B$_2$, respectively. To assess whether the material is in the clean or dirty limit, we evaluate the ratio of the BCS coherence length to the mean free path $\xi_0/l_0$, where $\xi_0$ can be found from:

\begin{equation}
    \xi_{eff}(0) =0.855\sqrt{\xi_0 l_0}
\end{equation}
This yields $\xi_0$ values of 243 nm for YRu$_3$B$_2$ and 998 nm for LuRu$_3$B$_2$, suggesting  that both systems fall in the dirty limit where $\xi_0 >> l_0$. The effective penetration depth $\lambda_{eff}(0)$can be determined from: 
\begin{equation}
    H_{c1}(0) = \frac{\Phi_0}{4\pi \lambda_{eff}(0)^2  } \ln(\frac{\lambda_{eff}(0)}{\xi_{eff}(0)})
\end{equation} 
This gives a $\lambda_{eff}(0)$ = 198.1 nm and 166.6 nm for YRu$_3$B$_2$ and LuRu$_3$B$_2$, respectively. These values are larger than the ideal London penetration depth $\lambda_L(0)$, which can occur, for example, due to disorder and non-local effects as well as from the simplified approximations inherent in both theoretical descriptions.
In addition, the analysis in Section \ref{SFWCalc} shows that, although quasi-flat bands exist in the band structure, the superfluid weight is mostly conventional; the geometric contribution is small because the dispersion of the bands at the Fermi level is large compared to the small superconducting order parameters for these low temperature superconductors.  


\section{Theory}
\subsection{Methods}
First-principles calculations were carried out within density functional theory (DFT) using the projector augmented-wave (PAW) method~\cite{Blochl1994ProjectorA,Kresse1999FromA}, as implemented in the Vienna \textit{ab initio} simulation package (VASP)~\cite{Kresse1996EfficiencyA,Kresse1996EfficientA}. The exchange-correlation potential was treated within the generalized gradient approximation (GGA) using the Perdew-Burke-Ernzerhof (PBE) functional~\cite{Perdew1996GeneralizedA}. A plane-wave energy cutoff of 500~eV was used, and the Brillouin zone (BZ) was sampled with a $\Gamma$-centered Monkhorst–Pack $\vb{k}$-point grid of $12 \times 12 \times 12$~\cite{Monkhorst1976SpecialA}. Irreducible representations of the electronic states were obtained using \textsc{irvsp}~\cite{Gao2021IrvspA}. Maximally localized Wannier functions (MLWFs) were constructed for \textit{R}-$s, d$, Ru-$s, d$, and B-$s, p$ orbitals in local basis~\cite{Mostofi2008Wannier90A,Mostofi2014An-updatedA,Pizzi2020Wannier90A,Deng2023Two-elementaryA,Jiang2025FeGeA}. Based on these MLWFs, the Fermi surfaces, density of states, and orbital-resolved band structures were computed using \textsc{WannierTools}~\cite{Wu2018WannierToolsA}.

Phonon spectra were calculated via the \textsc{Quantum ESPRESSO} package in the framework of density functional perturbation theory (DFPT). Electron-phonon coupling (EPC) properties were computed using the \textsc{epw} package~\cite{Giustino2007Electron-phononA, Noffsinger2010EPW:A, Ponce2016EPW:A, Lee2023Electron-phononA}, included within \textsc{Quantum ESPRESSO}~\cite{Giannozzi2009QUANTUMA, Giannozzi2017AdvancedA}. Electron-phonon matrix elements were first computed on coarse $\vb{k}$-point and $\vb{q}$-point meshes ($8\times8\times12$ and $4\times 4\times 4$) and then interpolated onto finer grids ($32 \times 32\times 32$) using maximally localized Wannier functions.

Superfluid-weight calculations were performed using the \textsc{Quantum ESPRESSO} package \cite{giannozzi_quantum_2009,giannozzi_advanced_2017}, together with an in-house developed code (not publicly available). We employed the generalized gradient approximation (GGA) with the Perdew-Burke-Ernzerhof (PBE) exchange-correlation functional \cite{perdew_generalized_1996} revised for solids (PBEsol) \cite{perdew_pbesol_2008}, in combination with the scalar relativistic optimized norm-conserving Vanderbilt pseudopotentials (ONCVPSP) from the PseudoDojo set \cite{van_setten_pseudodojo_2018}. We considered 25/11/16/3 valence electrons for Lu, Y, Ru, and B, respectively. Plane-wave energy cutoffs of 100 Ry for the wavefunctions and 1000 Ry for the density were chosen, along with a Methfessel-Paxton smearing \cite{methfessel_highprecision_1989} of 0.010 Ry. We computed the band-resolved superfluid weight with a $9\times 9\times14$ self-consistent $\vb{k}$-point grid with 100 points along each path segment in the non-self-consistent calculation. The total superfluid weight required a non-self-consistent $\vb{k}$-point grid of $63\times63\times98$ for convergence.

\subsection{Electronic band structure}
\label{ElPhCalc}
\begin{figure*}
    \centering
    \includegraphics[width=\linewidth]{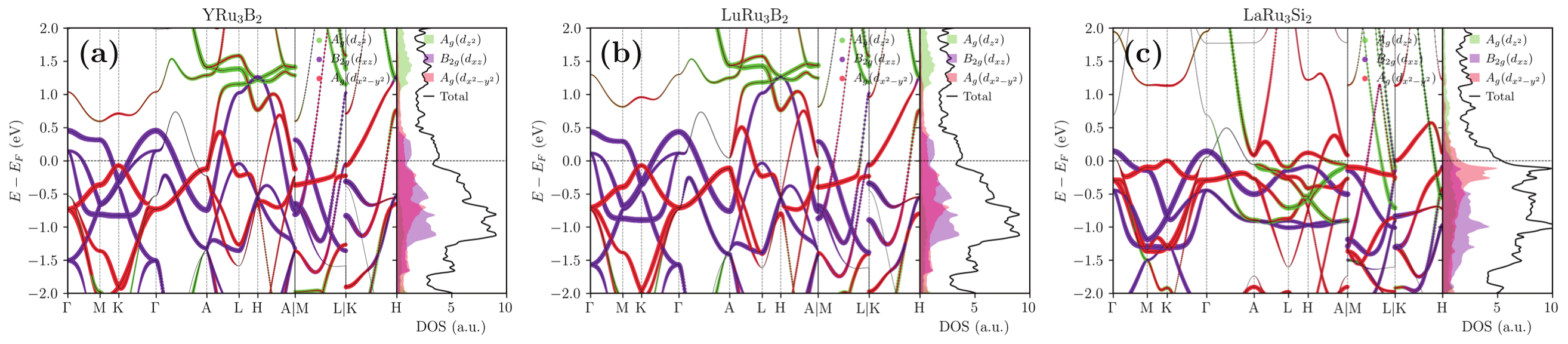}
    \caption{Orbital-projected band structure and density of states (DOS) of the $P6/mmm$ structure for (a) YRu$_3$B$_2$, (b) LuRu$_3$B$_2$, and (c) LaRu$_3$Si$_2$. The $d$ orbitals (specifically $d_{z^2}$, $d_{x^2-y^2}$, and $d_{xz}$) of Ru atoms are projected onto the band structure, with the radius of the circles proportional to the projected weight.}
    \label{fig:band-dos-all}    
\end{figure*}

To rationalize the suppression of $T_c$ in YRu$_3$B$_2$, and LuRu$_3$B$_2$ compared with LaRu$_3$Si$_2$ from an electronic-structure perspective, we compare their band structures  (see Fig.~\ref{fig:band-dos-all}). As the lattice constants contract from La to Y and Lu (Table~\ref{tab:lattice-constants-lambda}), the bands become more dispersive, most notably the quasi-flat band near $E_F$ from the Ru $d_{x^2-y^2}$ orbital (\textit{i.e.}, the red bands in Fig.~\ref{fig:band-dos-all} on the $k_z=0$ plane). In LaRu$_3$Si$_2$, this quasi-flat band spans $\sim 0.3$ eV \cite{DengLaRuSi2025}, whereas in YRu$_3$B$_2$ and LuRu$_3$B$_2$ its bandwidth increases to $\sim 0.7$ eV (Fig. \ref{fig:band-dos-all}). This broadening substantially reduces the DOS near $E_F$: the quasi-flat band contribution decreases from $1.589$ states/eV/spin in LaRu$_3$Si$_2$ to $0.754$~states/eV/spin in YRu$_3$B$_2$ and $0.753$~states/eV/spin in LuRu$_3$B$_2$, and the total DOS decreases from $5.308$~states/eV/spin in La to $3.541$~states/eV/spin in Y and $3.535$~states/eV/spin in Lu. In contrast, the Ru $d_{xz}$ DOS increases from $0.808$~states/eV/spin in La to $1.125$~states/eV/spin in Y and $1.123$~states/eV/spin in Lu (listed in \cref{tab:lattice-constants-lambda}). Here all orbital labels refer to the local coordinate system defined in Refs.~\cite{Deng2023Two-elementaryA,Jiang2025FeGeA,DengLaRuSi2025}. Since the EPC constant scales with the DOS as
\begin{equation}
    \lambda = 2 \frac{D(\mu)}{N}\frac{\hbar \ev{g^2}}{\hbar^2 \ev{\omega^2}},
    \label{eq:lambda_DOS}
\end{equation}
where $\ev*{g^2}$ is the Fermi surface averaged EPC strength, $N$ is the number of unit cell, $D(\mu)$ is the DOS at chemical potential $\mu$, and $\ev*{\omega^2}$ the McMillan average phonon frequency. The reduced $D(\mu)$ directly weakens EPC strength in YRu$_3$B$_2$ and LuRu$_3$B$_2$. This trend is consistent with recent experimental results on \ce{YRu3Si2}~\cite{kral2025discoveryhightemperaturechargeorder}, where replacing La by the smaller Y ion reduces the in-plane lattice constant $a$ from 5.715~\AA\ to 5.472~\AA\ and $T_c$ from $6.8$~K to $3.4$~K. Orbital-resolved band analysis further shows that, in \textit{R}Ru$_3$B$_2$ (\textit{R} = Y, Lu), the Ru $d_{x^2-y^2}$-derived quasi-flat band does not show an effective hole doping relative to LaRu$_3$Si$_2$. Instead, the Ru $d_{xz}$ and $d_{z^2}$ bands shift upward and become hole-doped (Fig.~\ref{fig:band-dos-all}). Moreover, the $d_{z^2}$-derived band lies substantially higher in energy (green bands in Fig.~\ref{fig:band-dos-all}), which weakens the $d_{z^2}$-bonding-driven structural instability toward the $Cccm$ (orthorhombically distorted) phase~\cite{max132distortion}.

\begin{table}[h]
    \centering
    \caption{Lattice constants (in unit of \AA) after full relaxation, total (and orbital-projected) DOS at the Fermi level (in unit of states/eV/spin), and electron-phonon coupling constant $\lambda_{e-ph}$ for YRu$_3$B$_2$, LuRu$_3$B$_2$, and LaRu$_3$Si$_2$.}
    \begin{tabular}{c|c|c|c}
        \hline\hline
        & YRu$_3$B$_2$ & LuRu$_3$B$_2$ & LaRu$_3$Si$_2$ \\
        \hline
        $a$ & $5.5037$ & $5.4762$ & $5.7154$ \\
        $c$ & $3.0267$ & $3.0144$ & $3.5755$ \\
        \hline
        $D(E_F)$ (total) & $3.535$ & $3.541$ & $5.308$ \\
        $D(E_F) (d_{x^2-y^2})$ & $0.754$ & $0.753$ & $1.589$ \\
        $D(E_F) (d_{xz})$ & $1.125$ & $1.123$ & $0.808$ \\
        $\lambda_{e-ph}$ & $0.477$ & $0.561$ & $0.831$ \\
        \hline\hline
    \end{tabular}
    \label{tab:lattice-constants-lambda}
\end{table}

\subsection{Phonons and electron-phonon coupling} \label{PhononsEPC}

\begin{figure*}
    \centering
    \includegraphics[width=\linewidth]{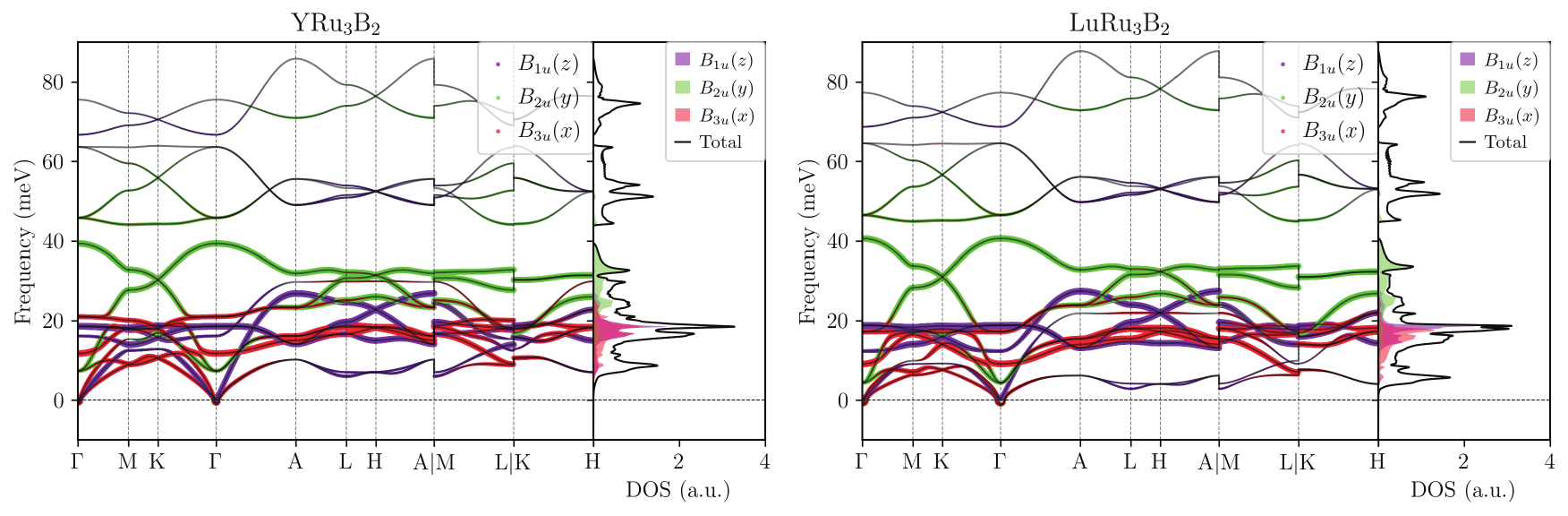}
    \caption{Mode-projected phonon band structure and density of states (DOS) of the $P6/mmm$ structure for YRu$_3$B$_2$ (left panel) and LuRu$_3$B$_2$ (right panel). The phonon modes (defined in local coordinates) of Ru atoms are projected onto the phonon band structure, with the radius of the circles proportional to the projected weight.}
    \label{fig:ph-band-orbital}
\end{figure*}

\begin{figure*}[htb]
    \centering
    \includegraphics[width=0.95\linewidth]{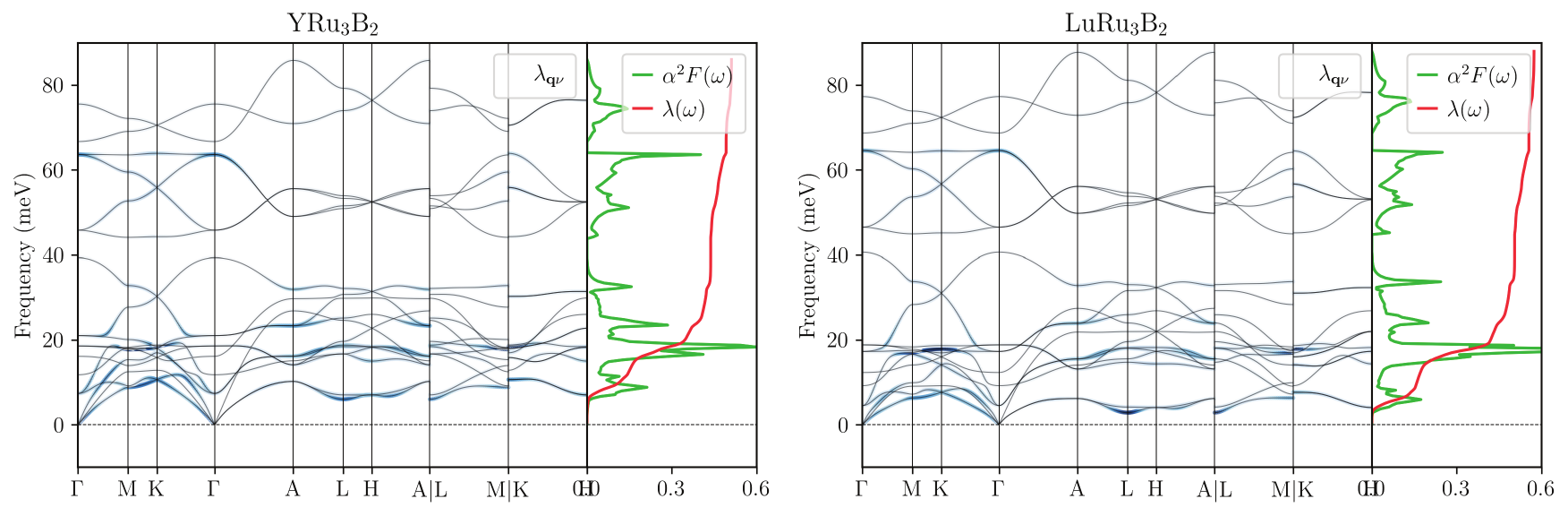}
    \caption{Phonon band structure, Eliashberg spectral function $\alpha^2F(\omega)$, and electron-phonon coupling constant $\lambda(\omega)$ for YRu$_3$B$_2$ (left) and LuRu$_3$B$_2$ (right).}
    \label{fig:phonons}
\end{figure*}

To assess phononic effects in \textit{R}Ru$_3$B$_2$ (\textit{R} = Y, Lu) and benchmark against LaRu$_3$Si$_2$, we compute phonon dispersions within DFPT and analyze their mode character and contribution to the electron-phonon coupling (EPC). Relative to LaRu$_3$Si$_2$, the phonon spectra of \textit{R}Ru$_3$B$_2$ (see \cref{fig:ph-band-orbital}) are overall shifted to higher energies, consistent with the smaller atomic mass of B with respect to Si, the smaller lattice constants, and stiffer effective force constants. This increases the characteristic phonon scale $\ev*{\omega^2}$ and therefore tends to reduce the EPC constant $\lambda$ for comparable $\ev*{g^2}$, particularly for the coupling between Ru $d_{x^2-y^2}$ states and Ru in-plane $x$ modes.

In LuRu$_3$B$_2$, we additionally identify a low-frequency branch at $\Gamma$ dominated by Lu in-plane ($xy$) motion that becomes soft (see Fig.~\ref{fig:ph-band-orbital}). The corresponding two-dimensional mode transforms as $\Gamma_{6}^{-}$ and is compatible with a symmetry lowering to $Amm2$ (No.~38) (the representation follows the convention on the  \textit{Bilbao Crystallographic Server}~\cite{aroyo2011crystallography, aroyo2006bilbao1, aroyo2006bilbao2}), \textit{i.e.}, a possible $C_3$-breaking distortion at low temperature. To avoid an artificial enhancement of $\lambda$ associated with an incipient instability within harmonic EPC theory, we evaluate EPC using the stable phonon spectrum obtained at relatively larger electronic smearing, yielding $\lambda \simeq 0.561$ (Fig.~\ref{fig:phonons}) and an estimated $T_c \simeq 3.27$~K from the Allen–Dynes modified McMillan equation~\cite{McMillan1968TransitionA,Dynes1972McMillantextquotesinglesA,Allen1975TransitionA} with $\mu^{*}=0.1$ for LuRu$_3$B$_2$. For YRu$_3$B$_2$, we obtain $\lambda \simeq 0.477$ and $T_c \simeq 1.88$~K. For reference, the $T_c\simeq 6.8$~K reported in Ref.~\cite{DengLaRuSi2025} was obtained using coarser $\vb{k}$ and $\vb{q}$-point meshes in a high-throughput workflow. The theoretical values of $\lambda$ with increased value for the smearing are close to the estimates based on experiments, namely \(\lambda^{exp}_{e-ph} \) = 0.44 (for YRu$_3$B$_2$) and 0.41 (for LuRu$_3$B$_2$) (see Table I in Results and Discussion.).
It is worth noting that matching the experimental $T_c$ within the Allen–Dynes framework requires $\mu^*\approx 0.14$ for Y and $\mu^*\approx 0.17$ for Lu.
The mode-projected phonon DOS and the phonon dispersions in Fig.~\ref{fig:ph-band-orbital} show that the Ru in-plane $x$ mode lies systematically below the corresponding $y$ modes, mirroring the trend in LaRu$_3$Si$_2$. As we show below and in \cref{fig:phonons}, this low-frequency Ru-$x$ branch provides the dominant contribution to $\lambda$.



To quantify the microscopic EPC, we project the \textit{ab initio} EPC tensor into a Wannier basis using \textsc{epw}~\cite{Giustino2007Electron-phononA, Noffsinger2010EPW:A, Ponce2016EPW:A, Lee2023Electron-phononA}. The leading real-space EPC matrix elements between Ru $d_{x^2-y^2}$ electrons and Ru $x$ phonons are essentially the same for both YRu$_3$B$_2$ (0.02024~Ry/Bohr), LuRu$_3$B$_2$ (0.02073~Ry/Bohr) and LaRu$_3$Si$_2$ (0.02043~Ry/Bohr). Given that the DOS at $E_F$ from Ru $d_{xz}$ and $d_{x^2-y^2}$ orbitals is comparable (\cref{tab:lattice-constants-lambda}), we also evaluate the corresponding EPC elements for Ru $d_{xz}$ electrons coupled to Ru $x$ phonons, obtaining 0.01254 (Y), 0.01231 (Lu), and 0.00925 (La). With a matrix element roughly half as large and similar DOS, \cref{eq:lambda_DOS} implies that the $d_{xz}$ channel contributes at the level of $\sim$1/4 of the $d_{x^2-y^2}$ channel, leaving the $d_{x^2-y^2}$ states as the dominant contributor to the total $\lambda$ as in LaRu$_3$Si$_2$~\cite{DengLaRuSi2025}.
Taken together, the near invariance of these real-space EPCs indicates that the reduced $\lambda$ in \textit{R}Ru$_3$B$_2$ originates from the suppressed $d_{x^2-y^2}$ DOS and the overall hardening of the phonon spectrum.

We further examine the EPC-driven superconducting properties by calculating and analyzing the superconducting gap function projected onto the Fermi surfaces. Using the \textsc{epw} package, we compute the anisotropic Eliashberg gaps for \ce{YRu3B2} and \ce{LuRu3B2} at $T=0.1$~K. 
The resulting momentum- and band-resolved gap function $\Delta_{n\vb{k}}$ exhibits a clear multigap structure: a dominant gap of $\sim 0.8$~meV opens on the Fermi-surface sheets with primarily Ru $d_{x^2-y^2}$ character (see Supplemental Material for details), whereas the Ru $d_{xz}$-derived sheets host a much smaller gap of $\sim 0.1$~meV. This separation of gap scales in \textit{R}Ru$_3$B$_2$ is similar to \ce{LaRu3Si2} and is consistent with a pairing mechanism controlled by the same mode-selective EPC channel~\cite{DengLaRuSi2025}.
Despite the similar gap anisotropy, the superconducting transition temperature is strongly reduced from $\sim 7$~K in \ce{LaRu3Si2} to $\sim 1$~K in \textit{R}Ru$_3$B$_2$. Within an EPC picture, the $T_c$ reduction is consistent with the broadening of the Ru $d_{x^2-y^2}$ quasi-flat band under lattice contraction, which lowers the DOS at $E_F$ and is expected to weaken $\lambda$ accordingly.
An instructive comparison can be made with the mean-field tight-binding pairing analysis (Section~\ref{MFtightbinding}). There, assuming the same pairing interaction for all orbitals, the DOS differences lead to a slightly larger gap on the $d_{xz}$ derived sheets than on the $d_{x^2-y^2}$ sheets. In contrast, our first principles Eliashberg results show that the $d_{x^2-y^2}$ gap dominates. This mismatch indicates that the pairing interaction is orbital dependent.

\subsection{Mean-field tight-binding model analysis of the superconducting order parameter structure} \label{MFtightbinding}

We construct a tight-binding model from DFT by generating Wannier functions for the two kagome compounds. Using this model, we perform self-consistent mean-field superconductivity calculations following the approach described in Ref.~\cite{DengLaRuSi2025}, with an on-site attractive interaction applied uniformly to each orbital. This calculation yields superconducting order parameters for all 32 orbitals, revealing the relative contribution of different atoms and orbitals to superconductivity. The interaction strength is chosen to produce superconducting gaps consistent with the experimental $T_c$. We verified that the qualitative trends described below remain robust across a range of interaction values.

In both YRu$_3$B$_2$ and LuRu$_3$B$_2$, we find that the three $d_{xz}$ orbitals of Ru have the same and the largest order parameter. Of the other contributions, the Ru $d_{x^2-y^2}$, $d_{yz}$, and $d_{xy}$ orbitals (three each) are within the same order of magnitude as the Ru $d_{xz}$ one. The contributions from the remaining Ru orbitals ($s$ and $d_{z^2})$, and all Lu and B orbitals, are one or more orders of magnitude smaller. Superconductivity is thus primarily caused by pairing related to the orbitals of the Ru atoms that form the kagome structure. Here, it is interesting to note that DFT analysis of superconductivity for LaRu$_3$Si$_2$ found the Ru $d_{x^2-y^2}$ orbital to also dominate the superconducting order parameter~\cite{DengLaRuSi2025}. However, the orbital content of the pairing gap differs because of the dominant density of states from $d_{x^2-y^2}$ close to Fermi level in LaRu$_3$Si$_2$, while in YRu$_3$B$_2$ and LuRu$_3$B$_2$ the $d_{xz}$ and $d_{x^2-y^2}$ contribute similarly to the density of states, as shown in \cref{fig:band-dos-all}. Our mean-field description reflects mainly the density of states, as the interaction is assumed uniform for all orbitals; for a more refined calculation, the orbital-dependence of interactions as well as possible nearest-neighbor interactions would be needed. The DFT calculations in Section~\ref{PhononsEPC} show that the gaps are different for different orbitals and the one of $d_{x^2-y^2}$ orbital is the largest, i.e., the density of states alone does not determine pairing.

\subsection{Superfluid weight and penetration depth calculations}
\label{SFWCalc}

We calculate the superfluid weight (superfluid stiffness) tensor $D_s = D_{geom} + D_{conv}$, including both the conventional contribution $D_{conv}$, proportional to the band dispersion, and the geometric one $D_{geom}$, arising from the quantum geometry of the bands~\cite{peotta:2015,Huhtinen2022}. 
To simplify the computations, we utilize the zero-temperature formula for the superfluid weight which assumes uniform pairing~\cite{Liang2017} ($i,j$ are the Cartesian coordinates $x,y,z$):
\begin{align}
    D_{conv}^{ij} &= \frac{1}{V} \sum_{ \mathbf{k}m }
    \frac{ \Delta^2 }{ \sqrt{\epsilon_m^{2} + \Delta^2}^{3}}
    \partial_i \epsilon_m \partial_j \epsilon_m 
    \label{eq:dconv-orig} \\
    D_{geom}^{ij} &= \frac{1}{V} \sum_{\mathbf{k}, m,n\neq m} 
\frac{ \epsilon_n - \epsilon_m }{ \epsilon_m + \epsilon_n } 
                    \left[ \frac{ \Delta^2}{ \sqrt{ \epsilon_m^2 + \Delta^2 }} - \frac{ \Delta^2}{ \sqrt{\epsilon_n^2 + \Delta^2 } } \right]    \nonumber \\
    & \times (\langle\partial_i m|n\rangle \langle n|\partial_j m \rangle + \text{H.c.}).
    \label{eq:dgeom-orig}
\end{align}
Here $ \epsilon_m$ is the energy dispersion shifted by the Fermi level, and it is a function of momentum $ \mathbf{k}$ with $m$ as the band index. The energy dispersions as well as the Bloch functions $|m\rangle$ and their derivatives are obtained by DFT. The superconducting gap $\Delta$ is estimated from the experimental $T_c$ via the BCS relation $\Delta \simeq 1.76~T_c$, which gives $\Delta = 0.106$~meV for the Y and $0.147$~meV for the Lu compound. The values for the superfluid weight components are summarized in \cref{tab:summary:sw}. Our calculations reveal that the superfluid weight in these compounds is dominated by the conventional contribution. This is expected given the dispersive nature of bands near the Fermi level and that quantum geometric contributions become significant only when the pairing gap is comparable to or exceeds the bandwidth, which is not the case here. The conventional contribution is essentially insensitive to the value of the gap, as it is dominated by the band dispersion near the Fermi level rather than the magnitude of the pairing scale. In contrast, the geometric contribution exhibits a strong dependence on the superconducting gap. Increasing the gap to 2~meV enhances the geometric term by approximately two orders of magnitude. Nevertheless, the geometric contribution remains negligible compared to the conventional term.


\begin{table*}[t]
\caption{Overview of total superfluid weight components computed from DFT data of YRu$_3$B$_2$ and LuRu$_3$B$_2$
with the uniform pairing assumption, and London penetration depths obtained from them. 
The calculation was done with $\Delta$ = 0.106 (Y) and 0.147~meV (Lu).}
\label{tab:summary:sw}
\centering
\begin{ruledtabular}

\begin{tabular}{@{\extracolsep{\fill}}lccccccccccc}

Compound
& xy $\lambda_L^{th}$
& z $\lambda_L^{th}$

& $D_s^{xx}$
& $D_s^{yy}$
& $D_s^{zz}$

& $D_{conv}^{xx}$
& $D_{conv}^{yy}$
& $D_{conv}^{zz}$

& $D_{geom}^{xx}$ 
& $D_{geom}^{yy}$
& $D_{geom}^{zz}$
\\
& (nm)
& (nm)
& (1/Hm)
& (1/Hm)
& (1/Hm)
& (1/Hm)
& (1/Hm)
& (1/Hm)
& (1/Hm)
& (1/Hm)
& (1/Hm)
\\
\hline
YRu$_3$B$_2$ & 32 & 24  
& 7.8 10$^{20}$ &  7.8 10$^{20}$ & 1.4 10$^{21}$ 
&  7.8 10$^{20}$ &  7.8 10$^{20}$ & 1.4 10$^{21}$ 
&  2.6 10$^{16}$ &  2.6 10$^{16}$ &  1.8 10$^{17}$\\ 
LuRu$_3$B$_2$ & 36 &  26
& 6.2 10$^{20}$ &  6.2 10$^{20}$ & 1.4 10$^{21}$ 
& 6.2 10$^{20}$ & 6.2 10$^{20}$ & 1.4 10$^{21}$ 
& 6.3 10$^{15}$ & 6.3 10$^{15}$ & 1.1 10$^{16}$\\ 
\end{tabular} 

\end{ruledtabular}

\end{table*}

\begin{figure*}
    \centering
    \includegraphics[width=\textwidth]{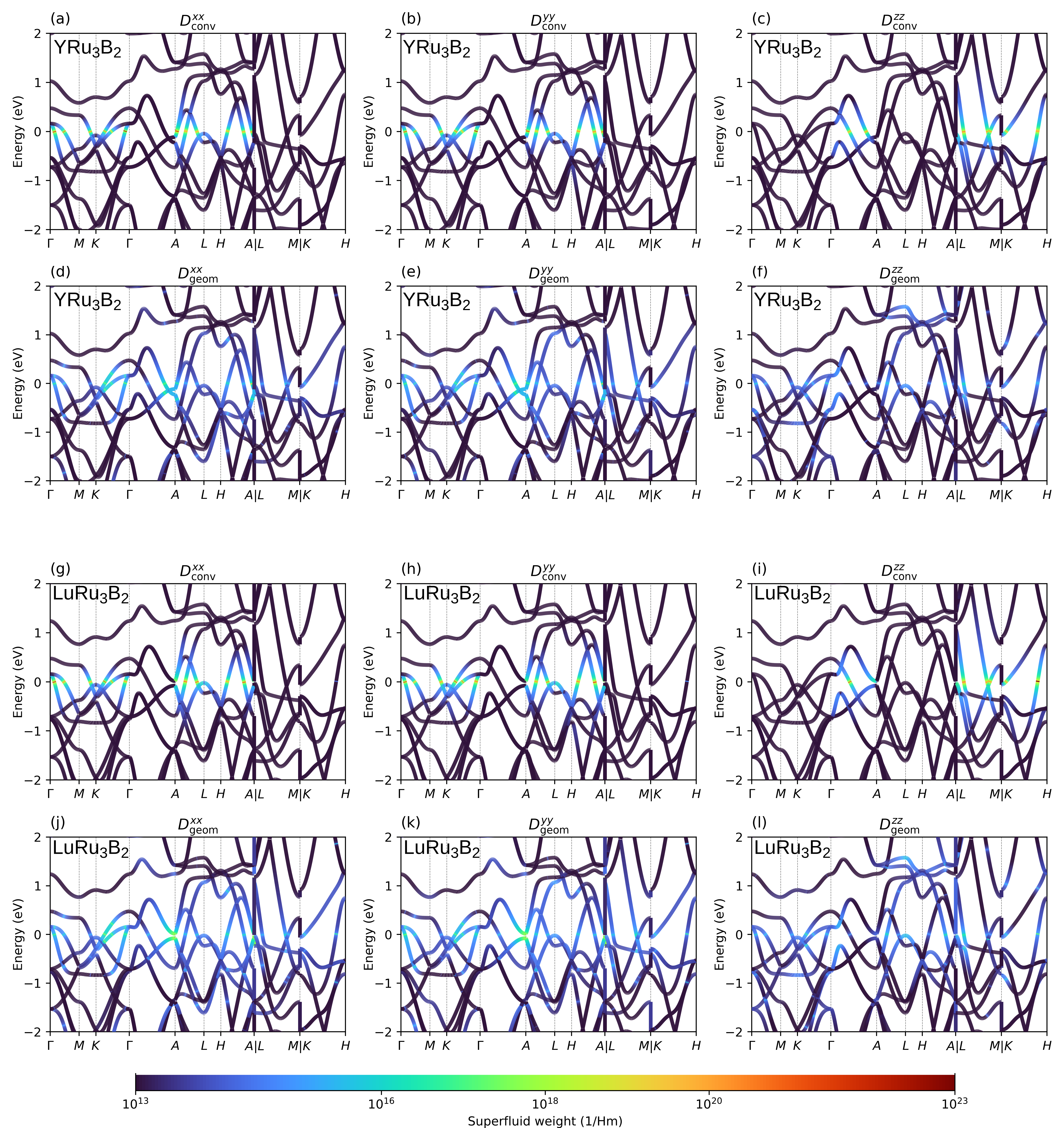}
    \caption{Integrands $D^{ij}(\mathbf{k},m)$ of the superfluid weight tensor components for YRu$_3$B$_2$ in the first two rows and LuRu$_3$B$_2$ in the last two, calculated using Equations (\ref{eq:dconv-orig})-(\ref{eq:dgeom-orig}) and DFT data, with $\Delta$ = 0.106 (Y) and 0.147~meV (Lu).} The conventional contribution is larger and concentrated on the Fermi energy while the geometric is smaller and more spread out. Note that the colorscale is logarithmic.
    \label{fig:Ds-high-sym-LuRu3B2}
\end{figure*}

To gain insight into which parts of the band structure contribute to the superfluid weight the most, we investigate the superfluid weight integrand $D_s^{ij}(\mathbf{k},m)$ when the total superfluid weight is $D_s^{ij}= \sum_{\mathbf{k},m} D_s^{ij}(\mathbf{k},m)$, see Equations (\ref{eq:dconv-orig})-(\ref{eq:dgeom-orig}). 
Figure \ref{fig:Ds-high-sym-LuRu3B2} shows the diagonal components of the superfluid weight integrand.
The largest contributions for the $xx$ and $yy$ of the superfluid‑weight integrand components are found along the $\Gamma-M-K-\Gamma$ and $A-L-H-A$ lines, while the $zz$ component vanishes on these planes. Conversely, $D_{\mathrm{conv}}^{zz}$ is non‑zero only along the $\Gamma-A$, $M-L$ and $K-H$ lines, where $D_{\mathrm{conv}}^{xx}$ and $D_{\mathrm{conv}}^{yy}$ vanish. This anisotropy follows from symmetry constraints and the fact that $D_{\mathrm{conv}}$ is directly proportional to the band dispersion. The crystal belongs to space group $P6/mmm$ (No. 191), which contains mirror planes perpendicular to $k_x$, $k_y$, and $k_z$ passing through $\Gamma$, as well as mirror planes containing the $A-L-H$ and $K-M-L-H$ paths set by the translational symmetries. These mirror symmetries force the component of the band dispersion normal to a mirror plane to vanish at mirror‑invariant points. Thus, the dispersion along $k_z$ is forbidden in $\Gamma-M-K$ and $A-L-H$ planes, while the $k_{x,y}$ one is suppressed along $\Gamma$-A, $M-L$ and $K-H$ lines.
To be precise, along the $H-K$ line, only the $xx$ component is forced to be zero by symmetry arguments. 
However, the zero dispersion along this direction, together with the other two non-dispersive directions related by the $C_{3z}$ symmetry, hinders the possibility of a large $yy$ dispersion, as shown in Figure~\ref{fig:Ds-high-sym-LuRu3B2}.

\begin{figure*}[h]
    \centering
    \includegraphics[width=0.9\linewidth]{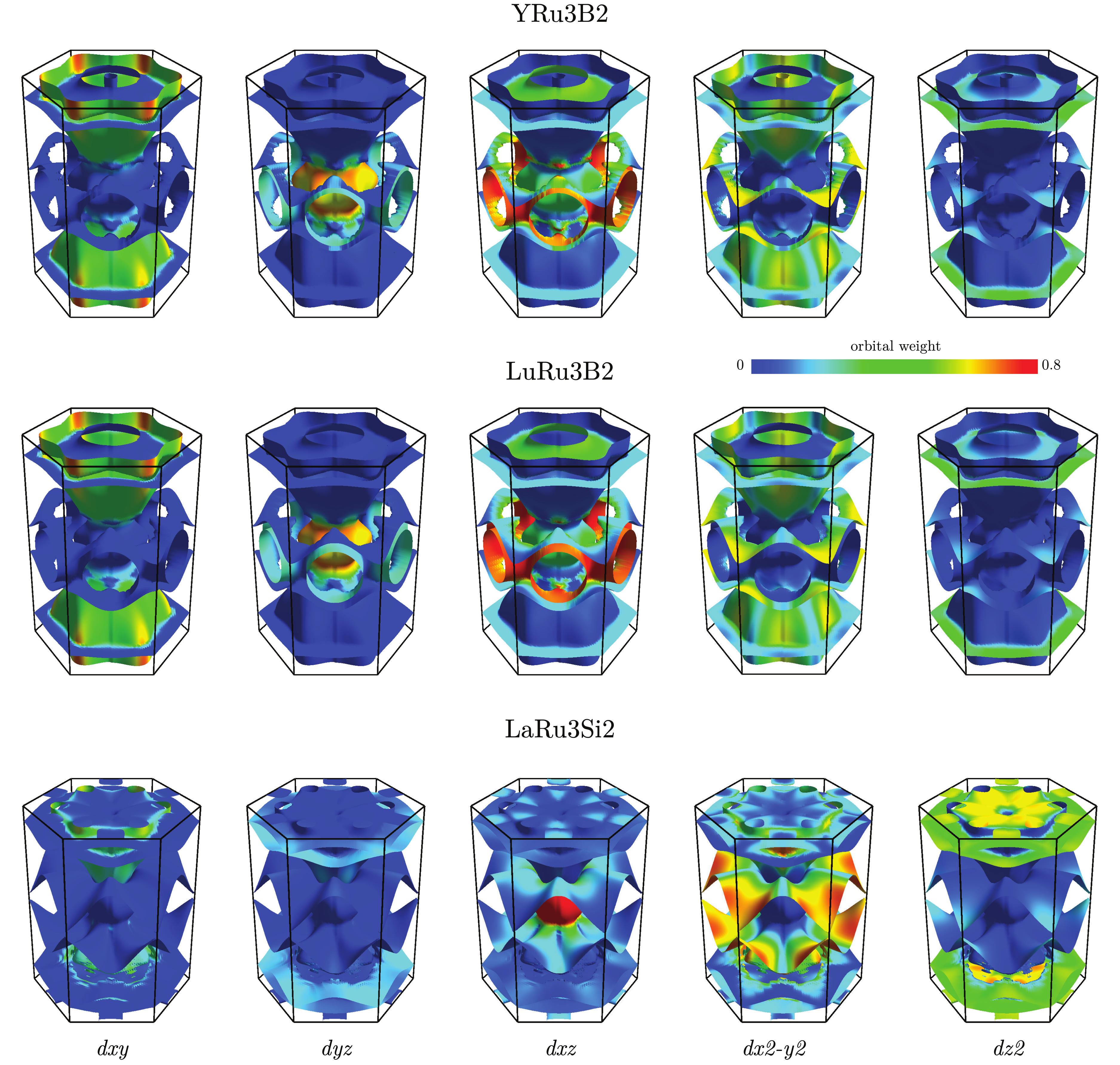}
    \caption{Orbital-resolved Fermi surfaces of YRu$_3$B$_2$, LuRu$_3$B$_2$, and LaRu$_3$Si$_2$ in $P6/mmm$ phase, with the color denoting the orbital weight of Ru $d$ orbitals in local coordinate.}
    \label{fig:fs-all}
\end{figure*}

The London penetration depth $\lambda_L$is related to the superfluid weight by 
\begin{equation}
    \lambda_L = (\mu_0 D_s)^{-1/2} .
    \label{eq: pen-depth}
\end{equation}
We use this formula with $D_s^{ii}$ to give the penetration depths, $\lambda^{th}_{z}$ in the $i=z$ direction, and $\lambda^{th}_{xy}$ in the $i=x,y$ directions which are nearly identical due to the approximate $xy$-isotropies of the system (see Figure \ref{fig:fs-all}), even when the system is not $xy$-symmetric. 
The penetration depth predictions are summarized in Table \ref{tab:summary}. The calculations 
show that YRu$_3$B$_2$ has slightly smaller penetration depths than LuRu$_3$B$_2$, however the error range partially overlaps. Both show a larger penetration depth in the \textit{xy} plane compared to the \textit{z} direction.




\section{Conclusions} \label{Conclusions}


We discovered two new superconductors, YRu$_3$B$_2$ and LuRu$_3$B$_2$, both predicted through ML-accelerated high-throughput screening of kagome lattice materials. These compounds crystallize in the hexagonal CeCo$_3$B$_2$-type structure with planar Ru kagome networks, and exhibit bulk superconductivity confirmed through magnetization, specific heat, and resistivity measurements with superconducting volume fractions close to 100\%.

The experimental realization of these materials validates the predictive power of combining ML with first-principles calculations for superconductor discovery, even though the observed critical temperatures are lower than initially predicted. Our theoretical analysis explains this discrepancy through the incipient phase transition leading to an excessive softening of phonon modes in the calculations.

The contraction of lattice parameters from LaRu$_3$Si$_2$ to the smaller Y and Lu analogs increases band dispersion, particularly for the quasi-flat kagome band derived from Ru $d_{x^2-y^2}$ orbitals. This enhanced dispersion reduces the density of states near the Fermi level, directly weakening the electron-phonon coupling constant and consequently suppressing $T_c$.
Detailed phonon and Wannier-resolved EPC analysis shows that the dominant coupling channel remains the Ru \(d_{x^2-y^2}\) states interacting with the low-frequency in-plane Ru-\(x\) branch, and its leading real-space matrix elements are nearly unchanged across the three compounds. The $d_{xz}$ and $d_{x^2-y^2}$ orbitals have similar densities of states near the Fermi level. This shift in orbital character reflects the band structure reorganization accompanying lattice contraction, with $d_{xz}$ and $d_{z^2}$ bands becoming hole-doped while the $d_{x^2-y^2}$ band remains near its original filling. Our mean-field analysis of the gaps, assuming uniform interaction for all orbitals, reflects the density of states and gives comparable gaps for both $d_{xz}$ and $d_{x^2-y^2}$ orbitals. However, the more microscopic DFT calculations reveal a two-gap behavior with the larger gap for the $d_{x^2-y^2}$ orbital. 

Our superfluid weight calculations demonstrate that conventional contributions dominate over quantum geometric effects in these materials. This is expected as quantum geometric contributions become significant only when the pairing gap is comparable to or exceeds the bandwidth, a condition not met in YRu$_3$B$_2$ and LuRu$_3$B$_2$. The calculated London penetration depths show moderate anisotropy between in-plane and out-of-plane directions, reflecting the quasi-two-dimensional character of the kagome structure.

The successful prediction-to-realization pipeline demonstrated here, from ML screening through first principles calculations to experimental synthesis and characterization, represents a significant advance in accelerating superconductor discovery. While the materials space of the 1:3:2 kagome family is manageable with current computational methods, this integrated approach becomes increasingly valuable when extended to larger materials spaces where exhaustive first-principles calculations are prohibitively expensive. The key is the strategic combination of rapid ML-based pre-screening to identify promising candidates, followed by targeted high-accuracy calculations and experimental investigation of the most viable predictions.

The broader significance of this work extends beyond these specific compounds. It provides a demonstration that the century-old paradigm of serendipitous superconductor discovery can be complemented, and in some cases replaced, by systematic, computation-guided materials design. As ML models improve and computational resources expand, this approach will become increasingly powerful for navigating the vast materials space and identifying superconductors with tailored properties. The successful realization of YRu$_3$B$_2$ and LuRu$_3$B$_2$ is a step toward this future, where theory, computation, and experiment work in concert to accelerate the discovery and characterization of quantum materials.

\begin{acknowledgments}
This work was supported by a collaboration between The Kavli Foundation, Klaus Tschira Stiftung, and Kevin Wells, and by the Jane and Aatos Erkko Foundation, the Keele Foundation and the Magnus Ehrnrooth Foundation, as part of the SuperC collaboration.
B.A.B, M.A.L.M., and P.T.
were supported by a grant from the Simons Foundation (SFI-MPS-NFS-00006741-01, B.A.B.; SFI-MPS-NFS-00006741-13,
M.A.L.M.; SFI-MPS-NFS-00006741-12, P.T.) in the
Simons Collaboration on New Frontiers in Superconductivity. This work is part of the Finnish Centre of Excellence in Quantum Materials (QMAT). We thank Kristjan Haule and Théo Cavignac for useful discussions.

\textit{Note added--}While preparing this manuscript, two works appeared on arxiv.org that report experimental observation of the YRu$_3$B$_2$ superconductor~\cite{Klimczuk2025,Hirschberger2025}.
\end{acknowledgments}

\newpage
\bibliography{YRu3B2_references}

\end{document}